%% file: paper.tex
\documentclass[10pt,conference]{IEEEtran}

\input{packages}
\input{macros}

\input{custom-macros-packages}

\usepackage{adjustbox} 
\usepackage{verbatim}
\newcommand{\eat}[1]{}

\begin{document}


\title{What Aspects of Mobile Ads Do Users Care About? \\An Empirical Study of Mobile In-app Ad Reviews}

\eat{
\author{\IEEEauthorblockN{Jiaping Gui, B, C and D}
\IEEEauthorblockA{University of Southern California\\
Los Angeles, California, USA\\
Email: \{jgui, b, c and d\}@usc.edu}
}}

\author{
\IEEEauthorblockN{Jiaping Gui\IEEEauthorrefmark{1},
Meiyappan Nagappan\IEEEauthorrefmark{2} and
William G. J. Halfond\IEEEauthorrefmark{1} }
\IEEEauthorblockA{\IEEEauthorrefmark{1}University of Southern
  California, Los Angeles, California, USA\\
 Email: \{jgui, halfond\}@usc.edu}
 \IEEEauthorblockA{\IEEEauthorrefmark{2}University of Waterloo, Waterloo, Ontario, Canada\\
Email: mei.nagappan@uwaterloo.ca}
}


\eat{
\numberofauthors{4}
\author{
\alignauthor
Jiaping Gui\\
       \affaddr{University of Southern California}\\
       \affaddr{Los Angeles, California, USA}\\
       \email{jgui@usc.edu}
\eat{       
\alignauthor 
author\\
       \affaddr{University of Southern California}\\
       \affaddr{Los Angeles, California, USA}\\
       \email{author@usc.edu}
\and
}
\alignauthor
Meiyappan Nagappan\\
       \affaddr{University of Waterloo}\\
       \affaddr{\vspace{10pt}Waterloo, Ontario, Canada}\\
       \email{mei.nagappan@uwaterloo.ca}
\alignauthor 
William G. J. Halfond\\
       \affaddr{University of Southern California}\\
       \affaddr{Los Angeles, California, USA}\\
       \email{halfond@usc.edu}
}
}

\maketitle

\input{Abstract} 




\input{Introduction} 
\input{ResearchQuestions} 
\input{Protocol} 
\input{Evaluation} 
\input{Threats} 
\input{Discussion} 
\input{Related} 
\input{Conclusion}
\input{References}

\end{document}

%% file: packages.tex
\usepackage{xspace}
\usepackage{multirow}
\usepackage{balance}
\usepackage{ctable}
\usepackage{listings}
\usepackage{color}
\usepackage{graphicx}
\usepackage{subcaption}
\usepackage{url}

\usepackage{algorithm}
\usepackage{algorithmicx}
\usepackage{algpseudocode}

\usepackage{amssymb}
\usepackage{amsmath}
\usepackage{amsfonts}

\usepackage{cleveref}
\usepackage{acronym}
\usepackage{fancybox}
\usepackage{verbatim}
\usepackage[normalem]{ulem}

%% file: macros.tex
\newcommand{\ie}{i.e.,\xspace}
\newcommand{\eg}{e.g.,\xspace}

\newlength{\MaxSizeOfLineNumbers}%
\definecolor{keywordcolor}{rgb}{0.8,0.1,0.5}
\definecolor{lightlightgray}{gray}{.96}
\definecolor{lightgray}{gray}{.925}
\definecolor{medlightgray}{gray}{0.7}
\definecolor{medgray}{gray}{0.4}
\definecolor{darkgray}{gray}{0.35}
\definecolor{nearblack}{gray}{0.15}

\crefname{component}{Component}{Components}

%% file: custom-macros-packages.tex
\newfloat{Program}{h}{eqn}

\definecolor{MyDarkGreen}{rgb}{0.0,0.4,0.0} 
\newcommand{\rqsec}[1]{\vspace{2pt}\noindent\textbf{#1}}

\newcommand{\adReviewNum}{529,827\xspace}

\newcommand{\rqma}{How frequent are ad related reviews among all app reviews?}
\newcommand{\rqmb}{What is the distribution of star ratings among ad reviews?}
\newcommand{\rqmc}{What are the common topics of ad complaint reviews?}

\newcommand{\rqt}[1]{\vspace{2pt}\noindent\textbf{#1}}

\newcounter{rqs}
\def\rqnumber{\addtocounter{rqs}{1}\arabic{rqs}}
\newcommand{\RQ}[1]{RQ\rqnumber: #1}

\acrodef{rq}[RQ]{research question}
\acrodef{ui}[UI]{user interface}
\acrodef{ct}[CT]{complaint topic}
\acrodef{mwu}[MWU]{Mann Whitney U}
\acrodef{uiat}[UIAT]{\textit{UIAutomator}}
\acrodef{man}[MAN]{mobile ad network}
\acrodef{lda}[LDA]{Latent Dirichlet Allocation}

\newcommand{\ctTitle}[1]{\textbf{#1}}

%% file: Abstract.tex
\begin{abstract}

In the mobile app ecosystem, developers receive ad revenue by placing
ads in their apps and releasing them for free. While there is evidence
that users do not like ads, we do not know what are the aspects of ads
that users dislike nor if they dislike certain aspects of ads more
than others. Therefore, in this paper, we analyzed the different
topics of ad related complaints from users. In order to do this, we
investigated app reviews that users gave for apps in the app store
that were about ads. We manually examined a random sample set of 400
ad reviews to identify ad complaint topics. We found that most ad complaints
were about \ac{ui} related topics and three topics were brought up the
most often: the frequency with which ads were displayed, the timing of
when ads were displayed, and the location of the displayed ads. Our
results provide actionable information to software developers
regarding the aspects of ads that are most likely to be complained about
by users in their reviews.

\end{abstract}

%% file: Introduction.tex
\section{Introduction}
\label{sec:introduction}

In just a matter of a few years, the global market has experienced a
tremendous increase in the number of apps that consumers use on their
smartphones. As of June 2016, both the Google Play and Apple app store
boasted over two million apps \cite{appstore2016}. Along with this
growth in apps, mobile advertising in apps has become an important
source of revenue for software developers
\cite{iab2013presentation}. In 2010, the mobile advertising industry's
revenue was just over half a billion dollars \cite{wsj2013mobileads}.
By 2018, analysts predict that revenue from mobile advertising will
reach 160 billion dollars  and account for 63\%
of all global digital advertising spending \cite{adspendingstat}.

In the mobile advertising ecosystem, there are four main stakeholders:
end users, developers, advertisers, and \acp{man}.  To earn ad
revenue, developers embed and display ads in their apps. \acp{man},
such as Google Mobile Ads and Apple iAD, facilitate the interaction
between developers and advertisers. To do this, \acp{man} maintain and
distribute libraries that enable developers to include ads served by
the \ac{man} in their apps.  When an end user clicks on or views ads,
the developer receives a small payment from the \ac{man} on behalf of
the advertiser.

An important additional, but somewhat indirect, player in the mobile
ad ecosystem is the app store.  Users can leave reviews and rate apps
in the app store.  This feedback can influence the behavior of other
users, who may avoid negatively rated apps, and can also be a source of
useful bug reports or suggestions for improvement for developers.
Prior studies have shown that these reviews cover a wide range of
topics, such as the app's functionality, quality, and performance
\cite{khalid2015mobile} and that specific areas of complaints, such as
user dissatisfaction with ads, can negatively impact the ratings an
app receives \cite{gui15icse}.  It is in the interest of developers to
avoid negative reviews and ratings as these will make their app less
appealing to new users or cause the app to be ranked lower by the app
store. In turn, fewer downloads of the app are likely to lead to fewer
ad clicks and views, which cuts into potential advertising revenue
that could be earned by developers.  However, in the case of ads, the
situation is more complicated.  Developers will not simply remove ads,
but must find a balance in their use of advertising that avoids a
negative experience, but still enables them to earn advertising
revenue.

The effect of reviews and ratings on developers' advertising revenue
motivates them to understand what aspects of ads could cause negative
experiences.  However, developers lack practical and even basic
information about which ad-related aspects are more or less likely to
produce a negative experience for their users.  Although many
developer blogs (\eg \cite{blogGuidance, blogGuidance1,
  blogGuidance2}) attempt to provide such guidance, and even ad
networks often suggest ``best practices'' (\eg \cite{officialGuidance,
  officialGuidance1}), this information is often anecdotal, lacks
rigorous evidence to support the advice, or is too generic to provide
developers with concrete guidance. Furthermore, developers lack a
systematic ability to analyze and understand ad related
reviews. Although there has been extensive study of app store reviews
(\eg \cite{chen2014ar, khalid2015mobile, mcilroy2016analyzing}), this
work has not focused on ad specific complaints.

To address this issue, we conducted an empirical analysis of ad
related reviews.  In this paper, we present the results of this
investigation, which enabled us to identify many different aspects of
ads that frequently trigger ad complaints. To carry out this
investigation, we performed a systematic approach as below:

\begin{enumerate}
\item We began by identifying ad related complaints from a corpus of
  over 40 million app store reviews.  We found that there were, in
  fact, a large number of user reviews that discussed mobile
  advertising.
\item We then analyzed the ratings and text of the reviews and found
  that those that mentioned advertising were disproportionately likely
  to receive lower ratings.
\item In a manual analysis, we analyzed a statistically significant
  sample of these reviews to identify the most common topics of end
  user complaints.  Our findings were that most complaints were
  related to how ads interfered or interacted with \ac{ui} related
  aspects of the mobile app.  In particular, we found that \ac{ui}
  issues relating to how frequently the ad was shown, when the ad was
  displayed, and where the ad was placed were the most frequently
  mentioned complaints. For non-visual aspects, the behaviors such as
  the ad automatically downloading files or changing system settings,
  blocking or crashing the host app's execution were the most
  frequently mentioned.
\end{enumerate}

Overall, these results showed clear trends in users' ad related
complaints that can help developers to better understand the aspects
they should be most concerned about when placing ads into their
apps. Better understanding of these aspects can improve the overall
app user experience while allowing developers to continue to take
advantage of the potential mobile ad revenue.

The rest of our paper is organized as follows. In Section \ref{sec:rqs}, we
introduce and motivate each of the research questions we address in
this paper.  In Section \ref{sec:protocol}, we describe the infrastructure
and protocol for collecting the ad meta data. Then in
Section \ref{sec:evaluation}, we describe the details and results of the
analysis we carried out for each of the research questions. The
threats to the validity of our results are discussed in
Section \ref{sec:threats}.  Next, in Section \ref{sec:discussion} we discuss the
implications of our findings and how these motivate future work in
different areas.  Finally, we cover related work in Section \ref{sec:related}
and summarize our findings in Section \ref{sec:conclusion}.

%% file: ResearchQuestions.tex
\section{Research Questions}
\label{sec:rqs}

Our investigation broadly focuses on end users' app store reviews that
relate to mobile advertising.  To better understand end users'
reactions to ads, we first focus on identifying how many such reviews
exist in the app stores, then we analyze the distribution of ratings
associated with these reviews, and finally we identify the ad related
topics in the reviews.  Below we more formally introduce and motivate
our \acp{rq}.

\rqt{\RQ{\rqma}}

An app store allows users to write textual reviews of the apps they
have downloaded.  If users have had a negative experience with mobile
advertising, they are likely to comment on this in their reviews.
Therefore, our first, and most basic, research question examines the
frequency of app reviews that include comments on mobile advertising.
The results of this \ac{rq} can serve to inform developers at to how
prevalent such reviews are in the corpus of all reviews.  We also are
interested in determining if this frequency varies by category. In
other words, are certain categories of apps more or less likely to get
comments related to ads?

\rqt{\RQ{\rqmb}}

Along with the textual review, the app store also allows users to
leave a numeric rating for each app.  This is typically provided by a
user on a scale of one to five, with five being the highest, and is
often referred to as ``awarding stars'' since the graphical
representation of a rating has traditionally been a star for each
rating point.  Developers care about these ratings because they
influence how app stores display apps in response to a user search.
Higher rated apps tend to be given more priority when displayed to the
user.  Therefore, in this research question we are interested in
determining the ratings distribution of ad related reviews.  We expect
that results for this RQ could indicate the type of influence ad
related reviews are having on an app's overall rating.

\rqt{\RQ{\rqmc}}

Complaints in ad related reviews may be due to numerous reasons.  For
example, users may be upset when an ad is interfering with the display of
important information or appearing too frequently. For developers it
is important to understand what aspects of ad usage or behavior is
causing user complaints so that they can focus their efforts on these
aspects.  Therefore, in this research question we are interested in
determining the ad complaint topics, that is the specific aspect or
issue related to ads that users are complaining about in a review. The
results of this RQ would inform developers as to the most problematic
aspects of ad usage and guide them in determining how to improve their
apps' ad usage.

%% file: Protocol.tex
\section{Data Collection}
\label{sec:protocol}

In this section, we describe briefly our protocol for obtaining the
data that is used to address each research question listed in
Section \ref{sec:rqs}.

To collect app reviews, we crawled the Google Play app store everyday
using a farm of systems for over two years (January 2014 to October
2016), to download every new release (\ie APK) of the app and its
associated meta-data, such as average user rating, user reviews (500
at a time), and their corresponding app ratings, among other
things. For the collection, we downloaded the top 400 ranked apps (as
ranked by Distimo \cite{distimo}) in 30 officially recognized app categories that
resulted in a corpus of more than 10,000 apps and over 40 million app
reviews.

%% file: Evaluation.tex
\setcounter{rqs}{0}

\section{Results and Discussion}
\label{sec:evaluation}

In this section, we discuss the details of the approach we employed to address each of the \acp{rq}, present the results we obtained, and discuss the implications of these results with respect to each of the \acp{rq}.

\input{adFrequency}
\input{adRatings}
\input{adTopics}

%% file: adFrequency.tex
\subsection{RQ1: \rqma}

\rqsec{Approach:} To answer this research question, we applied the
regular expression (\ie regex = ad/ads/advert*) to filter out ad
related reviews from all of the collected app reviews (see
Section \ref{sec:protocol}). We chose these particular keyword variations
based on guidelines provided in the related work that also manually
examined user reviews for different types of user complaints
\cite{khalid2015mobile}. We then counted the frequency of ad related
reviews and their percentage with respect to all of the reviews. We
also calculated similar information for each of the 30 standard app
categories.

\rqsec{Results:} We obtained in total \adReviewNum ad related reviews
out of over 40 million app reviews. This indicates that approximately
1\% of all user reviews dealt with in-app ads. The frequency of ad
reviews varies among the 30 app categories. Nine categories have an ad
review ratio over 1.5\%, and eight have less than 0.8\%. 
As for the absolute numbers, seven have over 30,000 ad
related reviews, and one has over 50,000. The median number and ratio
of ad reviews per category are 11,669 and 0.96\% (comparable to the
overall ratio), respectively.

\rqsec{Discussion:}


From the results, we can see that ad related reviews are
non-negligible. In fact, there are a large amount of such reviews from
end users. From this perspective, developers are able to extract
useful information about mobile ads to improve the user experience of
their apps, since these reviews are the feedback directly from end
users. This further motivates us to inspect ad reviews in the
following section to determine what ad aspects that users care about
and thus matter to developers. Ostensibly, the ratio of ad reviews is too small to cause developers' attention. Nonetheless, these reviews do have a measurable impact on the ratings, as we demonstrate in the next \ac{rq}. Past research has also shown the similar impact that ratings of apps could be affected when they receive poor reviews \cite{gui15icse}.
Moreover, mobile advertising has become one of the main
resources for the revenue developers receive when they publish free
apps in the app store. Hence, this ad ratio, albeit with a small percentage, is enough to be worthy of developers' attention.


We also found that the frequency of ad reviews varied by category. In
particular, two app categories, \textit{BRAIN} and
\textit{ENTERTAINMENT}, were high in both the number and ratio of ad
reviews. 
Another two
categories that are worthy to note were \textit{COMICS} and
\textit{MEDICAL}. They had a high ad review percentage in spite of
the relatively small number of ad reviews. This reflects that end
users commented on such kinds of apps not as actively as in other app
categories, but negative experiences with ads tended to be one of the
big problems that would cause users to complain.
Such a higher number or ratio in any of the four categories mentioned above could either mean that developers are more aggressively embedding ads in apps of these categories or that users of such apps have a lower tolerance for ads. In either case, developers need to be more cautious to prevent the loss of users and thereby sustained revenue.

%% file: adRatings.tex
\subsection{RQ2: \rqmb}

\begin{table}[]
\centering
\caption{Distribution of reviews with respect to their star ratings}
\label{tab:ratingStats}
\begin{tabular}{|@{}c@{}|c|c|c|@{}c@{}|@{}c@{}|}
\hline
Review rating & $\star$     & $\star$ $\star$    & $\star$ $\star$ $\star$   & $\star$ $\star$ $\star$ $\star$  & $\star$ $\star$ $\star$ $\star$ $\star$ \\ \hline
\% of Ad related reviews       & 33.29 & 13.21 & 14.51 & 17.1 & 21.89 \\ \hline
\% of Non-ad related reviews       & 12.12 & 4.51 & 7.27 & 15.11 & 60.98 \\ \hline
\end{tabular}
\end{table}

\rqsec{Approach:} To address this research question, we leveraged the
same approach as in the previous RQ, but calculated the frequency
distribution of both ad and non-ad review ratings.

\rqsec{Results:} Table \ref{tab:ratingStats} shows the distribution of all
reviews with different rating stars. We can see that, for the ad
related reviews, almost half (about 46\%) of the reviews are
complaints (\ie have one or two star ratings). In contrast, most of
the non-ad reviews have five star ratings with only about 17\% being
complaints.

\rqsec{Discussion:} These results show that reviews mentioning ad
related topics are disproportionately more likely to be a complaint
than non ad related reviews. Thus these reviews, with their
corresponding ad complaint topics, can convey valuable information
about what kind of ad aspects developers should address to improve
their app ratings. The high ratio of low ratings among ad related
reviews also suggests that ad related complaints can have a
significant negative impact on app ratings. Intuitively, such results
are unsurprising, but do provide motivation to further investigate and
understand which aspects of ads developers can address to improve
their apps.

One might wonder if since ad reviews together comprise only a little
more than 1\% of all of the reviews, are they likely to have an impact
on ratings that would register with the developer. Consider the case
where an app has 1,000 non ad related reviews and 11 ad related
reviews, all with four-star ratings.  If one of these ad reviews had
only a one-star rating, then the app's overall average rating would
drop to 3.997, a decrease of about 0.003 stars. This is a small
number, but related work has shown that even such small changes are
sufficient to change the rating based ranking of an app
\cite{gui15icse}, which in turn, could cause apps to be displayed in a
less favorable position in response to a user search.

%% file: adTopics.tex
\subsection{RQ3: \rqmc}

\rqsec{Approach:}

To answer this research question, we carried out a manual analysis of
a large subset of the ad related reviews to determine the most common
ad complaint topics.  In this study we focused on ad related
complaints and not the general topics relating to ads. To identify
complaints, we first filtered the corpus of \adReviewNum ad related
reviews to leave only those that had received a rating of two stars or
less.  Prior studies have shown that these reviews are primarily
negative or complaints~\cite{khalid2014prioritizing}.  Based on the
corpus size of 246,370 ad related complaint reviews, we randomly sampled 400
reviews to give us a 95\% confidence level with a 5\% confidence
interval.  This sample size ensures a high degree of confidence that
our categorization results would be indicative of the larger
population.

We then analyzed the 400 reviews to identify the ad related complaint
topics.  Our analysis involved two phases.  The first phase was a
high-level classification of whether the review was actually a
complaint.  Essentially, this allowed us to filter out any reviews
from our original set of 400 that were rated with one or two stars,
but did not actually complain about ads.  In the second phase, we
analyzed each of the ad related complaints to determine the topics
they mentioned.  For this analysis, we used a straightforward process
proposed in a prior work that also analyzed app review complaints
\cite{khalid2015mobile}.  We chose a manual analysis over an automated
analysis, such as sentiment analysis, because in our experience, the
automated analyses were less accurate at identifying new and distinct
complaint topics than a manual analysis. Our steps were as follows:

\begin{enumerate}
\item Examine each ad complaint to determine the topic of the complaint.
\item If no topic could be identified, classify complaint as
    \textit{non-descriptive}.
\item If topic was identified and the topic has been previously
  created, categorize the complaint as part of that topic.
\item If topic was identified and the topic was not previously
  defined, create a new topic and categorize the complaint with that
  topic.
\end{enumerate}

\rqsec{Results and Discussion:} 

\vspace{1ex}
\emph{1) High Level Category Distribution}
\vspace{1ex}

\begin{figure}[!tbp]
\center
\begin{subfigure}[H]{0.40\textwidth}
  \includegraphics[width=\linewidth]{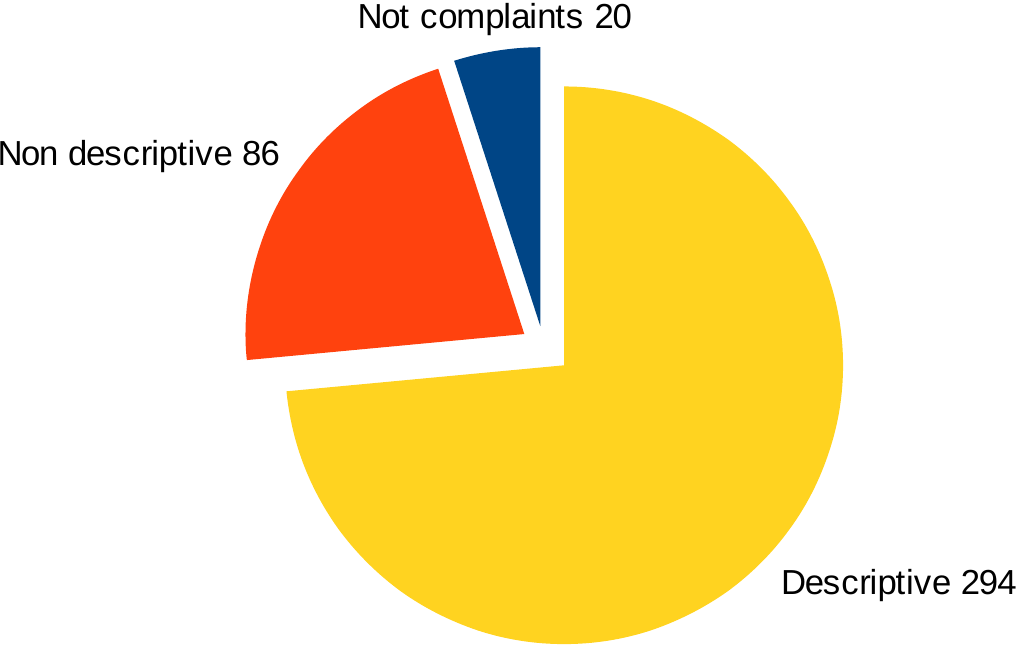}
\end{subfigure}%
\vspace*{2ex}
\caption{High level distribution of ad reviews.}
\label{fig:allStats}
\vspace*{-2ex}
\end{figure}

Our first result is a high-level classification of all 400 ad related
reviews.  This is shown in Figure \ref{fig:allStats}. In this figure, we
report the number of reviews that were not ad complaints
(``\textit{not complaints}''), the number of complaints for which a
topic could not be determined (``\textit{non-descriptive}''), and then
a broad categorization of the remaining complaints
(``\textit{descriptive''}).  At a high-level, we found that most of the ad
  reviews with low ratings were indeed complaints about ad aspects.
  Of the 400 reviews, 95\% (\textit{descriptive} + \textit{non
    descriptive}) had complaints, and only 5\% (\textit{non
    complaint}) were positive or neutral in their comments about the
  ads.

Among the two categories containing complaints, we found that about
20\% were \textit{non descriptive}.  Examples of comments we found in
this category were: \textit{`Ads are annoying'} and \textit{`Ads suck,
  suck'}. For such reviews, we could not extract useful information
related to their topic, except to determine that they were complaints
about ads.

\begin{figure}[!tbp]
\center
\begin{subfigure}[H]{0.24\textwidth}
  \includegraphics[width=\linewidth]{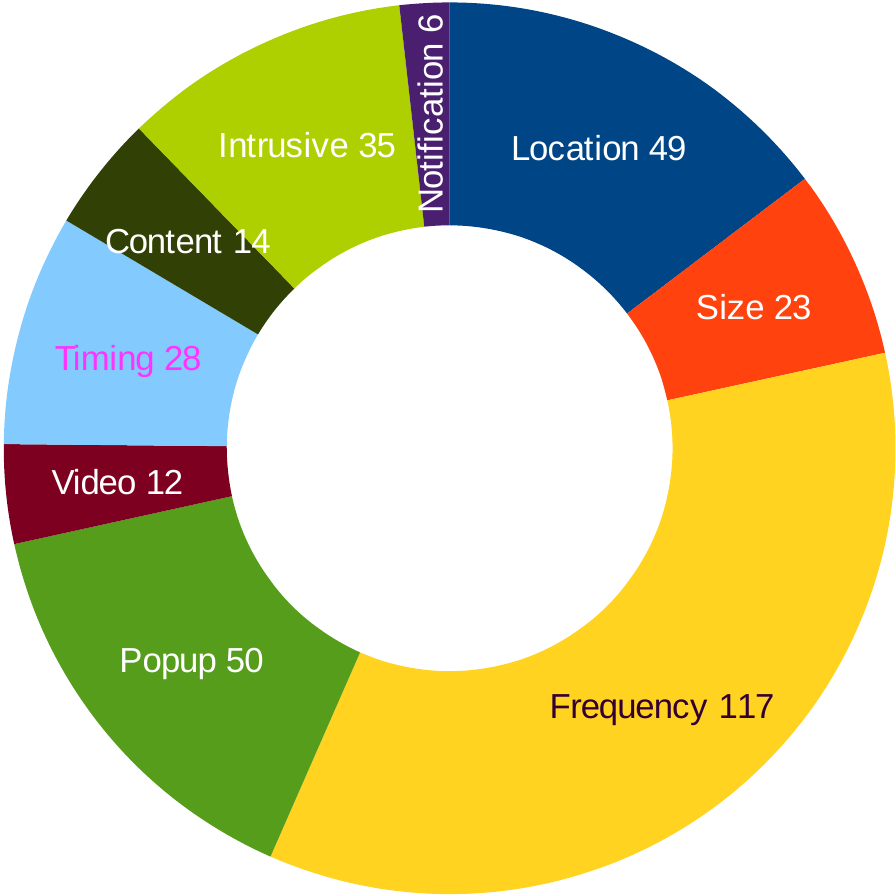}
  \caption{UI related.}
  \label{fig:uiStats}
\end{subfigure}%
\begin{subfigure}[H]{0.24\textwidth}
  \includegraphics[width=\linewidth]{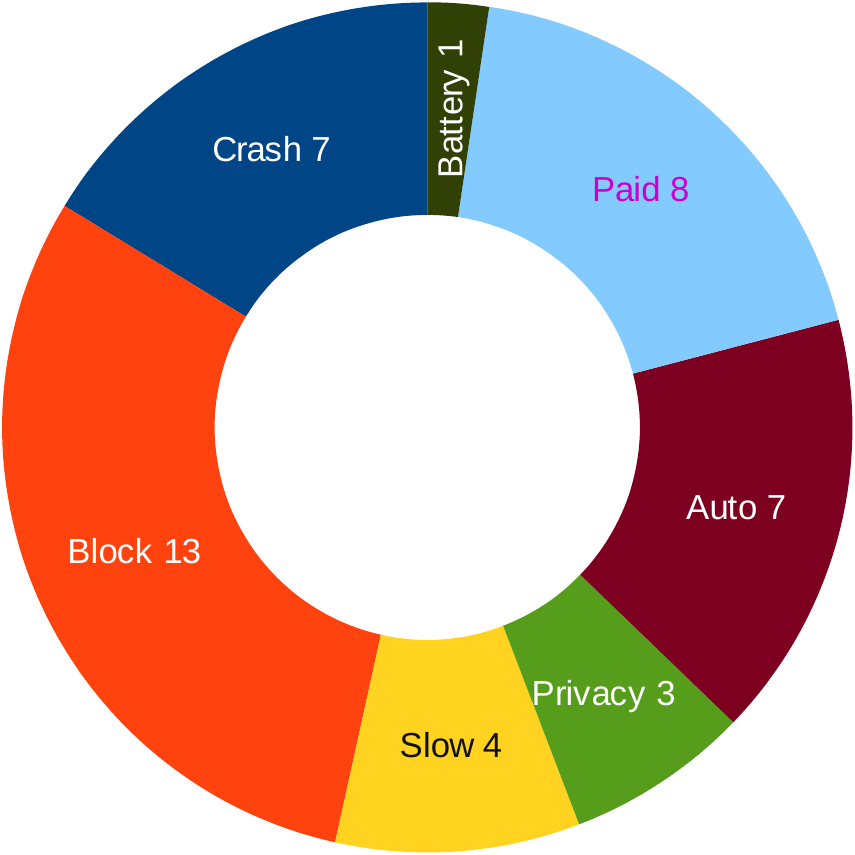}
  \caption{Non-UI related.}
  \label{fig:nonuiStats}
\end{subfigure}%
\caption{Topic distribution of descriptive ad complaints.}
\label{fig:pieStats}
\vspace*{-2ex}
\end{figure}

Of the remaining descriptive ad complaints, we found that most were
topics that could be considered \textit{\ac{ui} related}.  These
focused on visually observable aspects (\eg size and location) of the
ads and how they interacted or interfered with the \ac{ui}.
Altogether, reviews in these complaint topics represented about 66\%
of all of the reviews and their breakdown is shown in
Figure  \ref{fig:uiStats}.  The \textit{non-\ac{ui}} related ad complaint
topics dealt with ad functional properties (\eg a slow down in the
app's execution or unexpected audio).  They represented about 10\% of
all ad reviews and their breakdown is shown in
Figure \ref{fig:nonuiStats}. Note that the sum of the complaint topics in
Figure \ref{fig:pieStats} does not equal the 294 as shown in
Figure \ref{fig:allStats} since some reviews contained more than one
complaint. We now discuss the specific ad complaint topics that were
categorized as \textit{\ac{ui}}, \textit{non \ac{ui}} during our
manual analysis.

\vspace{1ex}
\emph{2) \ac{ui} Related Ad Complaint Topics}
\vspace{1ex}

The distribution of \ac{ui} related ad complaint topics is shown in
Figure \ref{fig:uiStats}.  Overall, we found nine distinct \ac{ui} related
complaint topics, each of which is shown in Figure \ref{fig:uiStats} along
with their frequency of occurrence in the reviews.  Below, we describe
the nine topics in more depth and give examples of users' complaints
about each topic.

\ctTitle{Interstitial (popup and video):} In general, there are two
types of ads that can be included in an app by a developer: banner and
interstitial. Ads that occupy the full screen are called interstitial
ads. Others, that appear as narrow horizontal strips, are
typically referred to as banner ads. Interstitial ads generally have a
higher payout than banner ads~\cite{interstitialpayout}. However,
users may react negatively to interstitial ads because they require
the user to view the ad for a time interval or click a close button to
return to the app.  For example, in a review of an app where an
interstitial ad popped up and interrupted a user's interaction with
the app, one user complained \textit{`...Ads pop up full
  screen. Uninstalled.'}. Another user complained \textit{`Ads have
  gotten out of control. Auto-play video ads with sound are not
  cool. Uninstalled.'}. This type of review represented 23.7\%
(62/262) of the total UI related complaints.

\ctTitle{Frequency:} This topic deals with how often ads appear
in an app. Current mobile ad networks pay developers based on the
number of ad clicks or ad impressions they achieve in their app.
Therefore, developers are incentivized to encourage user clicks and ad
views.  A developer may believe that one way to achieve this is by
displaying more ads in their app.  However, users may be annoyed by
too many ads, since it could be distracting or unsightly. In the UI
related ad reviews that we analyzed, we found many complaints (44.7\%:
117/262) were related to ad frequency. One user wrote \textit{`Ad
  overkill,Too many ads'} and another \textit{`Ads Ads Ads. As soon as
  I opened it, bang! Hit with an Amazon ad. That's enough for me,
  uninstalling now!!'}.

\ctTitle{Size:} This topic focused on how big an ad is in relation to
the app's \ac{ui}. To attract the user attention to ads, developers may be
tempted to make their ads bigger so they stand out.  However, as the
ad size becomes larger, it can affect the users' ability to interact
with the app. For example, \textit{`... ads at the bottom started
  getting bigger eventually blocking the next level button.'} and
\textit{`Stupid slot. Ads with slots covering the screen! People love
  casino on mobile but I think it is boring!'}. We found that 8.8\%
(23/262) of the ad UI related complaints were related to the size of
the ad.

\ctTitle{Location:} This topic includes complaints about where an ad
is within the app's \ac{ui}. Developers may place ads anywhere in the
\acp{ui} of their app.  Odd positions may increase the attention
brought to the ad, which could lead to more clicks.  However, ads in
certain positions, such as the middle of the page may be disruptive to
the usability of the app.  For example, one user complained
\textit{`Ads on the middle hate'} and another posted \textit{`Ads in
  the way. I can't read the quotes with the ads in my
  way. Uninstalled.'}.  Ad position directly impacts the layout of
other elements in the page. For instance, if an ad is displayed in the
middle of the screen, other elements can only occupy the upper half
and lower half of the page. In some cases, an ad may even overlap with
other visible elements. We found that 18.7\% (49/262) of the \ac{ui}
related reviews were related to the position of the ad in the screen.

\ctTitle{Notification:} Besides displaying relevant content, ads can
attract users' attention by sending an alert or notification to the
status bar. However, notifications can also trigger users'
complaints. A user commented that \textit{`Ads Notification is too
  bad'}. Another user graded the app as \textit{`Ads in your
  notification bar...  It is the most annoying thing ever!...'}. Out
of all \ac{ui} related complaints, six ad reviews (2.3\%) were about
ad notifications.

\ctTitle{Intrusive:} Another topic of ad \ac{ui} complaints is related
to whether ads interrupt the users' interaction with an app. For
example, if ads always popped up and needed a user's confirmation,
then the user could not play the app until confirming a yes or no
dialog. Ads affecting the user experience of the app incurred negative
reviews from users. A relatively large amount of ad \ac{ui} reviews
(13.4\%: 35/262) complained about the intrusive ads. Following are two
example complaints: \textit{`Ads are just too
  invasive. Uninstalled...'}, \textit{`Ads interfere with use. The ads
  are intrusive and take over the app making it unusable. Its not
  possible to try it out to see if you would like to purchase it.'}.

\ctTitle{Content:} This topic deals with what is in the ads shown to
users. Appealing ad content can help to catch users' attention.
However, we found evidence in the reviews to suggest a problematic
aspect to this.  Namely, users would complain when the content of an
ad repeated or did not change.  For example: \textit{`... I do however
  wish they weren't the same ads every single time ...'}, \textit{`Ad
  spam. The game is fine, but after every play i get spammed by the
  same video ad over and over. When i close one exactly the same one
  comes up after'}. We found that 5.3\% (14/262) of the \ac{ui}
related reviews that we examined were related to the content of ads.

\ctTitle{Timing:} A sure-fire way to make a user see an ad is to place
the ad on a landing page that is shown to the user when the app starts
or before the next level in a game app.  However, complaints from
users indicate that this may affect users' impressions of the app. For
example: \textit{`Ad helish!! Just work to carry stupid ads after
  launching. And locks your screen without any
  interaction. Annoyance.'} and \textit{`Ads Ads Ads. As soon as I
  opened it, bang! Hit with an Amazon ad. That's enough for me,
  uninstalling now!!'} were two of the reviews that we found regarding
this practice. We found that 10.7\% (28/262) of \ac{ui} related
complaints pertained to having ads appear at an undesirable time.

\vspace{1ex}
\emph{3) Non \ac{ui} Related Ad Complaint Topics}
\vspace{1ex}

The distribution of non \ac{ui} related ad complaint topics is shown
in Figure \ref{fig:nonuiStats}.  Overall, we found seven distinct non
\ac{ui} related complaint topics, each of which is shown in
Figure \ref{fig:nonuiStats} along with their frequency of occurrence in the
reviews.  Below, we describe the seven topics in more depth and give
examples of users' complaints about each topic.

\ctTitle{Blocking:} This topic includes ads that disabled the normal
functioning of an app.  For example, when the ad is running, the
execution of the host app is blocked and the end user cannot access
the app's primary functionality until the ad is interacted with by the
user. These reviews were the most frequent in the non \ac{ui} category
and represented 30.2\% (13/42) of non \ac{ui} ad complaints. For
example, one user stated that \textit{`Ads make it impossible to
  use. Ads refuse to play, so videos won't load. Spent ten minutes
  switching between a few videos and couldn't watch any.'}.

\ctTitle{Paid:} Besides publishing a `free but with-ads' version of apps,
developers usually provide paid apps that charge users for ad free
functions. However, if these paid apps do not behave as expected and
contain ads, this can cause user complaints. For example, a user
complained that \textit{`Ads in paid version. Donated to remove ads,
  but they still show up. Emailed support, got no response...'}. There
were in total eight reviews (19\%: 8/43) about this topic.

\ctTitle{Auto:} For the purpose of executing some functions, in-app
ads may automatically download files from remote servers or run in a
special manner (\eg turning on the audio and playing a downloaded
audio file even though the device is in a muted state). Such behaviors
can be considered malicious or extremely annoying. Once noticed, these
ads incite complaints from users. Following is an instance,
\textit{`Ads are truly annoying. I am really getting sick of the
  random auto launching of Playstore from within a game to suggest I
  download another game. Disappointing that a great game is being
  ruined by lousy marketing tactics'}. In the review results, about
(16.3\%: 7/43) of non \ac{ui} complaints were related to this topic.

\ctTitle{Crash:} Sometimes ads are poorly implemented and are not
compatible with app functionality. The end result is that the app
keeps crashing. In this case, users cannot interact with any function
of the app. There were several such complaints (16.3\%: 7/43) in the
result, such as \textit{`ad you put on crashes the game. Have to
  uninstall too many times. Fix the problem Or Stop using ad'}.

\ctTitle{Slow:} The inclusion of mobile ads can slow the functionality
of the app. The running of ads requires system resources, such as CPU
and memory, which are limited on the mobile device. As a result, less
resources can be allocated for the running of the host app, and this
slows down the app's execution. There were several reviews (9.3\%:
4/43) complaining about this aspect. One example user complained that
\textit{`Ads are irritating... Otherwise good app, Can anyone tell me
  how to block advertises?  It's little bit slow too while operating..
  Otherwise it would have been five stars.'}.

\ctTitle{Privacy:} In the mobile advertising ecosystem, ads inherit
permissions from the host app. This allows ads to access sensitive
information on the smartphone if the host app is granted the
permission. However, users may be upset to learn that in-app ads
obtain these permissions as well. Some examples include: \textit{`Ad
  supported apps suck. I absolutely refuse to install a free or ad
  supported app on my phone. The privacy intrusion and resource
  overload is unacceptable...'} and \textit{`Ads and permission. Ads
  and permissions granted are wrong'}. There were several complaints
(7\%: 3/43) about this topic.

\ctTitle{Battery:} Mobile in-app ads consume extra resources on the
system, such as energy. Components, such as display and network, are
two of the most energy consuming components on a mobile device
\cite{li14icsme, li2014making, wan15icst, li16icse}. These two
components also serve an important role in the mobile ad ecosystem
since they are used to retrieve and display ads. Even though energy
consuming behavior is not directly linked to the app or the ads it
contains, mobile ads routinely consume a significant amount of energy
and extreme levels of resource consumption may trigger user
complaints~\cite{gui15icse}.  There was one such complaint on this
topic (2.3\%: 1/43): \textit{`Ads. Ads everywhere. Popups
  also. Expected battery drain'}.

\begin{table}[]
\centering
\caption{Large-scale analysis of ad reviews of apps in each category
  (total of 30) to identify ad complaint topics. We report the number of
  categories in which each ad complaint topic occurs among the top five.}
\label{tab:large-scale-topics}
\begin{tabular}{|c|c|cc}
\hline
UI topics    & \# app categories & \multicolumn{1}{c|}{Non-UI topics} & \multicolumn{1}{c|}{\# app categories} \\ \hline
frequency    & 30                & \multicolumn{1}{c|}{paid}          & \multicolumn{1}{c|}{30}                \\ \hline
timing       & 30                & \multicolumn{1}{c|}{auto}          & \multicolumn{1}{c|}{30}                \\ \hline
location     & 28                & \multicolumn{1}{c|}{block}         & \multicolumn{1}{c|}{29}                \\ \hline
intrusive    & 24                & \multicolumn{1}{c|}{crash}         & \multicolumn{1}{c|}{29}                \\ \hline
content      & 12                & \multicolumn{1}{c|}{slow}          & \multicolumn{1}{c|}{28}                \\ \hline
popup        & 8                 & \multicolumn{1}{c|}{battery}       & \multicolumn{1}{c|}{1}                 \\ \hline
blank        & 6                 &                                    &                                        \\ \cline{1-2}
notification & 5                 &                                    &                                        \\ \cline{1-2}
size         & 4                 &                                    &                                        \\ \cline{1-2}
\end{tabular}
\end{table}

\rqsec{Additional Discussion:} Our analysis of the 400 randomly
selected reviews yielded sixteen distinct complaint topics.  Despite
being unable to similarly analyze the larger corpus of ad related
complaints, we were interested in identifying if there were possibly
additional topics that were undetected by our manual analysis.  To
investigate this, we used automated NLP based techniques to examine
the larger corpus for clusters of ad complaint topics and compared
this to our originally identified set.

To carry out this investigation, we applied the Word2Vec clustering
technique to the textual contents of the ad complaint corpus.  We used
Word2Vec because in our comparison of NLP techniques, it performed
most strongly at generating meaningful clusters (See
Section \ref{sec:threats}).  Prior to running the clustering technique, we
preprocessed the reviews through the following procedure: tokenizing
words, removing useless symbols and stopwords (\eg the, and, look),
and stemming words. The output of running the NLP technique was a list
of clusters, each of which contained a set of keywords that summarized
the cluster.  For each cluster, we calculated the relative word
frequency and then used this to identify complaint topics associated
with them and compared these topics to those identified by the manual
categorization.

The clusters very closely matched the manual categorization with two
exceptions.  We found one cluster associated with complaints related
to ads that appeared as blank space (\textit{blank}) and another
cluster related to ads whose display was tied to users' participation
in some sort of promotion activity (\textit{promotion}). The first of
these scenarios could occur when there is a change in the underlying
ad unit ID (a unique ID generated by an ad network that identifies the
app requesting ads to an ad network), the app's included ad library
has become obsolete, or there are connection problems between the app
and \ac{man}.  The second scenario seems to occur when users
participated in an activity with the purpose of obtaining an ad free
experience, but the app still displayed ads in some capacity.
Overall, the results of this larger scale indicate that the manually
identified ad complaint topics are highly representative of the larger
corpus of ad complaint reviews.

We would also like to confirm if the most popular ad complaint topics
are consistent with their manual counterparts.
Table \ref{tab:large-scale-topics} shows the top five ad complaint topics
for each app category after the large-scale analysis.  From the
results, we can see that for \ac{ui} related topics,
\textit{frequency}, \textit{timing}, and \textit{location}
are the most common ad related complaint topics and
are among the top three ad complaint topics for almost all of the 30
app categories.  Meanwhile \textit{paid}, \textit{auto},
\textit{block}, \textit{crash}, and \textit{slow} are the most common
ad non-\ac{ui} related complaint topics. This finding has the same
trends as our manual categorization results.

%% file: Threats.tex
\section{Threats to Validity}
\label{sec:threats}

\textbf{External Validity:} The analysis in our study was based on
reviews for only Android apps, so the results and conclusions may not
generalize to other platform apps (\eg iOS and Windows based), which
also represent a significant portion of the app marketplace. However,
we expect that since the underlying mechanisms of ad display are
similar, we would see similar feedback from end users.  In fact,
developers implement several versions of an app that can be published
on different platforms. These versions share the same or similar
functionality. In other words, the user experience of different
versions of the app in most cases is comparable and the differences
for end users are likely minor with respect to ads. Hence we argue
that mobile in-app ads impact the user experience similarly across all
platforms.

\textbf{Internal Validity:} We manually categorized ad reviews to
obtain ad complaint topics. This process may be biased by human error or
subjectivity and thus lead to incorrect tagging. To address this
threat, we revisited each ad review several times after all reviews
were initially categorized. In particular, each review was inspected
at least three times. The other authors randomly inspected ad reviews
to check the correctness of the categorization.

In addition, to confirm if the problems associated with ad complaints
existed in their corresponding apps, we further conducted a
qualitative study to validate these complaints. Such a study also
enabled us to make an accurate categorization for reviews with vague
or borderline descriptions. To do this, we installed and manually
interacted with the apps corresponding to each of 400 ad related
reviews. The apps that we interacted with were those for which we had
access to the app's primary functions as real users and could interact
with long enough to ensure several ad reload cycles. We registered as
a new user, if needed, before entering the main page. Once the ad
aspects complained about by end users were confirmed in the app, we
terminated the interaction. To ensure the ad functionality was fully
loaded, each app was interacted with for at least 5 minutes unless the
ad complaint was confirmed before that. The mobile device we used was
a Samsung Galaxy SII smartphone with a rooted Android 4.3 operating
system that is compatible with the original version of the APK
file. We focused on the UI related features since they are more robust
against the outside interference as compared to non-\ac{ui} aspects. Our
results showed that over 80\% of ad complaints were confirmed for
those apps that had ads displayed during the interaction. In other
words, the problems associated with most ad complaints in the reviews
exist in the corresponding apps. This further validates the
conclusions of our analysis for each of the \acp{rq}.

\textbf{Construct Validity:} We applied Word2Vec to cluster ad
complaint topics in each app category at a large scale. However, there may be
other NLP techniques that are more suitable for such analysis. To
mitigate the threat due to the choice of the NLP technique, we used
the identified ad complaint topics in the manual analysis to evaluate the
performance of three up-to-date NLP techniques. These were Word2Vec \cite{word2vec},
K-means \cite{kmeans}, and \ac{lda} \cite{lda}, all popular in the area of text
understanding. Our results showed that Word2Vec identified the most
complaint topics (15 out of the total 16), while K-means and \ac{lda}
identified 9 and 12 complaint topics respectively. Although it is possible
that other NLP techniques could identify even more, the goal of this
study was not to identify the best possible NLP technique but to
identify a reasonable one to use for the purposes of our \acp{rq}.


%% file: Discussion.tex
\section{Directions for Future Work}
\label{sec:discussion}

An important question to address in this paper is: what comes next?
To answer this question we conducted an investigation to learn about
how developers of the apps in our study have responded to their ad
complaints over time.  We began by downloading the latest versions of
our subject apps (latest as of December 2016). In total, we were able
to retrieve 322 of the 377 apps that corresponded to our reviews.  For
each of the apps for which we had been able to confirm the ad
complaint, we installed the latest version of them on the same device we used to confirm
the ad complaint and then interacted with it in the same way to
attempt to reproduce the underlying ad problem.  For each app, we
recorded if the problem was reproducible in the latest version.  We
only focused on the apps that had confirmed \ac{ui} related complaints
since these could be easily confirmed by visual inspection.

We found that a significant majority of the apps still had the
original ad related issue.  More specifically, we found that 87\% of
the apps were unfixed, while for 13\% of the apps we were unable to
find the reported issue.  Anecdotally, the fixed apps were typically
highly popular apps (\eg \textit{AngryBirds}) with a large base of
users, while the unfixed apps were typically much less popular.

These results raise interesting questions and motivate future work.
In particular, we were struck by the fact that for so many apps, a
significant and impactful topic of complaint had not been addressed.
We hypothesize two possible explanations for this.  First, developers
may be unaware of the impact or the significance of ad related
complaints.  This hypothesis motivates further investigations into the
impact of ad related complaints and the development and dissemination
of guidelines that can be inferred from the complaints.  Second,
developers may be aware of the ad complaints, but believe they cannot
change the app as that would lead to an unacceptable reduction in ad
revenue.  This hypothesis motivates further investigations into
analyses that can help developers quantify the tradeoffs between
maximizing mobile ad revenue and minimizing negative user experiences
with ads.

%% file: Related.tex
\section{Related Work}
\label{sec:related}

In previous work \cite{gui15icse}, Gui and colleagues conducted an
empirical evaluation on quantifying different costs of mobile
ads. Among these costs, one was the impact of ads on app ratings. In
that study, they observed that 4\% of all complaints in app reviews
were about ads. However, they did not categorize ad complaint topics or
validate the ad related complaints in the corresponding apps. Ruiz and
colleagues~\cite{ruiz2014impact} examined the impact of ad libraries
on ratings of Android mobile apps, and found that integrating certain
ad libraries could negatively impact an app's rating. While their work
focuses on the ad library at the app level, ours is more fine grained
and looks at the specific topics of complaints for ad related reviews.

A large amount of related studies have focused on user reviews and
ratings at the app level. Palomba and colleagues
\cite{palomba2015user} tracked how applications addressing user
reviews increased their success in terms of rating. Specifically, they
monitored the extent to which developers accommodated crowd requests
and follow-up user reactions as reflected in their ratings. Galvis
Carre{\~n}o and colleagues \cite{galvis2013analysis} relied on
adapting information retrieval techniques to automatically extract
topics from user comments. These topics were useful during the
evolution of software requirements. AR-Miner \cite{chen2014ar} was
proposed to analyze the content of user reviews. It was able to
discern informative reviews, group and rank them in order of
importance. Villarroel and colleagues \cite{villarroel2016release}
took a step further to design CLAP, an approach to automatically
categorize user reviews into suggestions, cluster related reviews, and
prioritize the clusters of reviews to be implemented in the next app
release. Khalid and colleagues \cite{khalid2015mobile} studied user
reviews from 20 iOS apps where they uncovered 12 types of user
complaints. The most frequent complaints they found were functional
errors, feature requests, and app crashes. ARdoc
\cite{panichella2016ardoc} and SURF \cite{di2016would} were proposed
to summarize user reviews by classifying useful sentences. Iacob and
colleagues designed \cite{iacob2013retrieving} MARA, a prototype for
automatic retrieval of mobile app feature requests from online
reviews, and relied on \ac{lda} to identify common topics across
feature requests. Guzman and colleagues \cite{guzman2014users} also
used \ac{lda} to group fine-grained app features in the reviews.  All
of the above described work targets app level reviews, and does not
directly categorize ad related reviews.

Another group of related work investigated ad fraud detection in
mobile apps. PUMA \cite{hao2014puma} is a programmable framework that
separates the logic for exploring app execution from the logic for
analyzing app properties. One of its analyses was for ad fraud
detection that identified small, intrusive, and too many ads per
page. Similarly, DECAF \cite{liu2014decaf} was designed and
implemented to detect various ad layout frauds for Windows Store
apps. Crussell and colleagues~\cite{crussell2014madfraud} developed an
analysis tool, MAdFraud, which automatically ran many apps
simultaneously in emulators to trigger and expose ad fraud by
analyzing HTTP requests. In contrast to their work, we are not
studying ad fraud or its impact, but rather examining ad related
reviews and extracting ad complaint topics.

Previous studies of mobile ads have also been conducted from different
perspectives. Gui and colleagues \cite{gui16greens} proposed several
lightweight statistical approaches to measure and estimate mobile ad
energy consumption. Eprof \cite{pathak2012energy} was presented as a
fine-grained energy profiler for smartphone apps and was one of the
first work to evaluate the energy consumption of third-party ad
modules. Ruiz and colleagues \cite{ruiz2014ad} carried out a broad
empirical study on ad library updates in Android apps. The results
showed that ad library updates were frequent, and suggested
substantial additional effort for developers to maintain ad
libraries. Li and colleagues \cite{li2016investigation} investigated
the use of common libraries in Android apps, and collected from these
apps 240 libraries for advertisement. Liu and colleagues
\cite{liu2015efficient} explored efficient methods to de-escalate
privileges for ad libraries in mobile apps. The system they developed
contained a novel machine classifier for detecting ad
libraries. Rasmussen and colleagues \cite{rasmussen2014green} analyzed
the effects of advertisement blocking methods on energy
consumption. None of these studies were concentrated on ad related
reviews and ratings.

Another group of related work conducted surveys and proposed different
methods or models to identify factors that influence consumers'
responses to mobile ads. Leppaniemi and
colleagues~\cite{leppaniemi2005factors} investigated factors, such as
the marketing role of the mobile medium, development of technology,
one-to-one marketing, and regulatory issues, which influence the
acceptance of mobile advertising from both industrial and consumer
points of view. With this information they built a conceptual model of
consumers' willingness to accept mobile advertising. Blanco and
colleagues~\cite{blanco2010entertainment} suggested entertainment and
informativeness as precursory factors of successful mobile advertising
messages, after an empirical study using structural modeling
techniques. Henley and colleagues~\cite{hanley2011smartphone}
conducted a study to investigate college student exposure to and
acceptance of mobile advertising in the context of an integrated
marketing communication strategy, and found that incentives were a key
motivating factor for advertising acceptance. In contrast to our
study, these approaches are based on users' response to mobile ads
through surveys and focus on the psychology behind ad acceptance. 
We consider such studies to be complementary to our focus on identifying
the top ad complaint topics.

%% file: Conclusion.tex
\section{Conclusions}
\label{sec:conclusion}

Currently, millions of smartphone users download free apps from app
stores and developers receive ad revenue by placing ads in these
apps. In fact, ad revenue has become one of the most important sources
for software developers to compensate for the cost of an app's
development. In this paper, we carry out experiments on a large scale
dataset of mobile advertising to investigate what kind of ad aspects
are complained about the most. We found that users complain about ad
visual aspects more often than the non-visual aspects.  We also found
that in many app categories, the most common ad complaint topics are
similar. Intuitively, more exposure of mobile ads to end users helps
improve the chance of ad impressions/clicks that increase the ad
revenue. But improper exposure is detrimental to the user experience
of an app which in turn negatively impacts the ad revenue developers
receive. App developers thus should carefully make a trade off to
maximize their ad revenue.

Based on our study in this paper, we suggest that when
developers design ad \ac{ui} during the implementation, it will
benefit them if they accommodate the following three criteria:

\begin{enumerate}
\item ad display timing: improper time or long duration
  (especially video ads) of ad display could cause a negative user
  experience;
\item ad display: some pages like the landing page may not be user friendly to display ads. The high frequency of ads among different pages is likely to distract the user's attention, and thus result in end user complaints;
\item visual layout: displaying ads in visually obstructive locations (\eg middle
  in the screen or close to clickable buttons) could interfere with
  the user's interaction with the app.
\end{enumerate}

When embedding ads into apps, developers should also pay attention to
ad non-\ac{ui} functions. In particular, it is not a good idea to
display ads in the so-called paid version of an app, since this has a
direct conflict with the expectations of users. Blocking app level
functionality to focus attention on ads is another design decision
that causes complaints by end users.  Furthermore implementation
decisions or lack of adequate testing that lead to the app crashing or
slowing down the app's running could also negatively impact the user
experience with the app.

Our work suggests multiple areas for future work.  In particular, we
plan to correlate different ad aspects to app ratings so as to
understand their relationship and identify more specific best
practices with respect to mobile ads. We would also like to carry out
controlled experiments and surveys that allow developers to determine
the impact of their ad related choices on user ratings.

%% file: References.tex

\balance
{
\bibliographystyle{IEEEtran} 

\bibliography{softwarequalitylab,ad}
}